\documentclass[prl,twocolumn]{revtex4}
\usepackage{graphicx}
\usepackage{dcolumn}
\usepackage{bm}
\usepackage{amsmath}
\usepackage{amsfonts}

\newcommand{\bra}[1]{\langle #1|}
\newcommand{\ket}[1]{|#1\rangle}

\newcommand{\D}[2]{D_{\mathbf{#1,#2}}}
\newcommand{\ba}[2][]{b^{#1}_{\mathbf{#2}}}
\newcommand{\bc}[2][]{b^{\dagger\,#1}_{\mathbf{#2}}}
\newcommand{\ma}[2][]{a^{#1}_{\mathbf{#2}}}
\newcommand{\mc}[2][]{a^{\dagger\,#1}_{\mathbf{#2}}}
\newcommand{\pa}[3][]{p^{#1}_{#2,\mathbf{#3}}}
\newcommand{\pc}[3][]{p^{\dagger\,#1}_{#2,\mathbf{#3}}}
\newcommand{\ca}[2]{c_{#1,\mathbf{#2}}}
\newcommand{\cc}[2]{c^{\dagger}_{#1, \mathbf{#2}}}
\newcommand{\sumk}[1]{\sum_{\mathbf{#1}}}

\newcommand{\da}[3][]{d_{#1 \mathbf{#2},#3}}
\newcommand{\dc}[3][]{d^{\dagger}_{#1 \mathbf{#2},#3}}

\newcommand{\al}[2]{\alpha_{#1,\mathbf{#2}}}
\newcommand{\be}[2]{\beta_{#1,\mathbf{#2}}}

\newcommand{\bec}[2]{\bar{\beta}_{#1,\mathbf{#2}}}

\newcommand{\K}[4]{\mathcal{K}^{#1,#3}_{#2,#4}}

\newcommand{\comm}[2]{\lbrack #1,#2\rbrack}

\begin{document}
\title{Stimulated scattering and lasing of intersubband cavity polaritons}
 
\author{Simone \surname{De Liberato}$^{1,2}$}
\author{Cristiano Ciuti$^{1}$}
\affiliation{$^1$Laboratoire Mat\'eriaux et Ph\'enom\`enes Quantiques, Universit\'e Paris  Diderot-Paris 7 and CNRS, UMR 7162, 75013 Paris, France} \affiliation{$^2$Laboratoire Pierre Aigrain,
\'Ecole Normale Sup\'erieure and CNRS, 24 rue Lhomond, 75005
Paris, France}

\begin{abstract}
We present a microscopic theory describing the stimulated scattering of intersubband polariton excitations in a microcavity-embedded two-dimensional electron gas. In particular, we consider the polariton scattering induced by the spontaneous emission of optical phonons.  Our theory demonstrates the possibility of final-state stimulation for the scattering of such composite excitations, accounting for the deviations from perfect bosonicity occurring at high excitation densities.  By using realistic parameters for a GaAs semiconductor system, we predict how to achieve a quantum degenerate regime, leading to ultralow threshold lasing without population inversion.
\end{abstract}
\maketitle

The scattering of  bosons from an initial to a final state can be stimulated, i.e., enhanced by the occupation of the final state. This remarkable property is in stark contrast with the behavior of fermions, such as electrons, whose scattering is Pauli blocked by final state occupation.
In low-energy matter there are no elementary bosons, yet composite particles acting like bosons can be obtained when an even number of fermions are bound together, such as atoms containing an even total number of nucleons plus electrons. In condensed matter systems\cite{Leggett}, the attractive interaction between two electrons can give rise to bosonic particles. Examples are Cooper pairs of electrons in metallic superconductors or Coulomb bound electron-hole pairs (excitons) in semiconductors. The strong coupling of an exciton with a microcavity photon produces the so-called exciton-polariton states, whose very small mass (inherited from the photon component) favors quantum degeneracy and the onset of stimulated scattering, responsible for exciton-polariton lasers\cite{exc_las} emitting in the near infrared.
  
\begin{figure}[t!]
\begin{center}
\includegraphics[width=9.5cm]{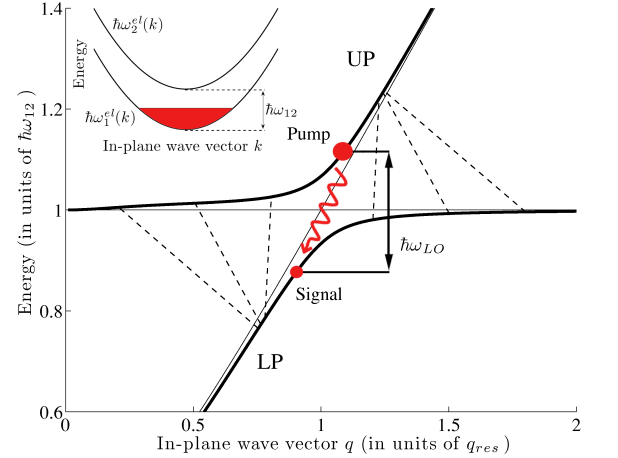}
\caption{\label{fig1}A typical energy dispersion (in units of the intersubband transition energy $\hbar \omega_{12}$) of intersubband cavity polaritons versus in-plane wave vector (in units of the resonant wave vector $q_{res}$). Due to the interaction with bulk optical phonons, a polariton initially pumped in the upper polariton (UP) branch  can scatter into a final state (signal mode) in the lower polariton (LP) branch by emitting an optical phonon with energy $\hbar \omega_{LO}$ ($36$ meV for GaAs). The considered modes have Hopfield coefficients $\beta_{UP,\mathbf{q'}} = \beta_{LP,\mathbf{q}}= 0.5$. The dashed lines indicate the same kind of scattering process by changing the in-plane momentum of the initial state along the upper polariton branch. Inset: the energy dispersion of the quantum well electronic conduction subbands versus electron wave vector $k$. In the ground state, a dense electron gas populates the first conduction subband.}
\end{center}
\end{figure}

Recently a novel kind of cavity polariton excitations has been discovered in a microcavity-embedded two-dimensional electron gas, \cite{Dini_PRL} and an intense research activity is currently expanding
\cite{Ciuti_vacuum,SST,Aji_APL,Ciuti_PRA,Luca_APL,Simone_PRL,Aji_excited,Luca_PRL,EL_model}. 
These light-matter excitations are the result of the strong coupling between a microcavity photon mode and the transition between two conduction subbands of the doped quantum well system.  A intersubband excitation has a well definite resonance frequency simply because the quantum well conduction subbands have parallel energy-momentum dispersions (see Fig. \ref{fig1}). In contrast to Cooper pairs or excitons, intersubband excitations do not correspond to any bound state originated from an attractive fermion-fermion interaction. This explains the remarkable robustness of intersubband cavity polaritons even at room temperature and the possibility of accurately tailoring their properties just by tuning the size of the quantum well or the density of the two-dimensional electron gas in the fundamental subband.  Even if intersubband excitations do not correspond to any bound electronic states, {\it a priori} this fact does not preclude the occurrence of stimulated scattering. Yet, to the best of our knowledge, a theory for the stimulation phenomena involving these composite excitations is still missing. Apart from the fundamental interest, a comprehensive understanding of the stimulated scattering of intersubband cavity polaritons  could pave the way to the remarkable realization of ultraefficient semiconductor lasers emitting in the mid and far infrared.

In this Letter, we present a microscopic theory of the stimulated scattering of intersubband cavity polariton excitations of a two-dimensional electron gas. In particular, we will consider the polariton scattering induced by the coupling with optical phonons, which is typically the most important interaction affecting semiconductor intersubband transitions, while Coulomb interactions are known to produce only moderate renormalization effects \cite{Imamoglu}.
Starting from the fermionic Hamiltonian for the quantum well electronic system and by using an exact iterative commutation procedure,  we are able to determine the phonon-induced
polariton scattering for an arbitrary number of excitations in the initial and final intersubband cavity polariton modes. Our results indeed prove the possibility of final-state stimulation of the intersubband cavity polariton scattering. 
Our theory also provides exactly the deviations from perfect bosonicity, occurring at high excitation densities. 
We apply our results to the case of a GaAs system with realistic losses and consider the case of intersubband cavity polariton lasing under resonant optical pumping.

We consider the Hamiltonian  $H=H_{lm}+H_{phon}$
where $H_{lm}$ is the light-matter term for the microcavity-embedded quantum well system, while $H_{phon}$ describes the dynamics of bulk optical phonons  and their the coupling to the quantum well electrons. Namely:
\begin{widetext}
\begin{eqnarray}
\label{H}
H_{lm}&=&\sum_{\mathbf{k},j=1,2}\hbar\omega^{el}_{j,k}\cc{j}{k}\ca{j}{k}+\sumk{q}\hbar\omega_{c, q}\mc{q}\ma{q}
+\sumk{k,q} \hbar\chi(q)\mc{q}\cc{1}{k}\ca{2}{k+q}+ \hbar\chi(q)\ma{q}\cc{2}{k+q}\ca{1}{k},\\
H_{phon}&=&\sum_{\mathbf{q},q_z} \hbar\omega_{LO,\mathbf{q},q_z}\dc{q}{q_z}\da{q}{q_z}+
\sum_{\substack{\mathbf{k,q},q_z \\ i,j=1,2}
}\hbar C_{ij,\mathbf{q},q_z} \da{q}{q_z}\cc{i}{k+q}\ca{j}{k}+\hbar C_{ij,\mathbf{q},q_z}\dc{q}{q_z}\cc{j}{k}\ca{i}{k+q},
\nonumber
\end{eqnarray}
\end{widetext}
where $\ca{j}{k}^\dagger$, $a_{\mathbf{q}}^\dagger$ and $\da{q}{q_z}^\dagger$ are the creation operators  respectively for an electron in the quantum well conduction subband $j$ with in-plane wave vector $\mathbf{k}$, a cavity photon with in-plane wave vector $\mathbf{q}$ and an optical phonon with three-dimensional wave vector $(\mathbf{q},q_z)$. Their respective energies are $\hbar\omega^{el}_{j,k}$, $\hbar\omega_{c,q}$ and $\hbar\omega_{LO,\mathbf{q},q_z}$ (the wave vector dependence of the optical phonon energy is negligible) and their phases are chosen in order to make
the coupling coefficients $\chi(q)$ and $C_{ij,\mathbf{q},q_z}$ real.
Being all the interactions spin conserving, we omit the spin degree of freedom for the electrons.
The photon polarization is meant to be Transverse Magnetic (TM) in accordance with the selection rules of quantum well intersubband transitions. Note that,  neglecting the conduction band non-parabolicity, the second subband dispersion is such that $\omega^{el}_{2,k} = \omega^{el}_{1,k} + \omega_{12}$, as depicted in the inset of Fig.1. Moreover, for typical photonic wave vectors $\mathbf{q}$, we can safely approximate $\omega^{el}_{j,\lvert \mathbf{k+q} \lvert}\simeq \omega^{el}_{j,k}$. 
The light-matter Hamiltonian $H_{lm}$ is then diagonalized by introducing the polariton creation operators
\begin{eqnarray}
\label{pol}
p_{\eta,q}^\dagger=\al{\eta}{q} a_q^\dagger+\be{\eta}{q} b_q^\dagger
\end{eqnarray}
where $\eta=\{LP,UP\}$ denotes the polariton branch index, $\al{\eta}{q}$ and $\be{\eta}{q}$ are real Hopfield coefficients describing the light and matter component respectively, while $\hbar \omega_{\eta,q}$ are their corresponding energies (see Fig. 1).  $\bc{q}$  is given by the expression \begin{eqnarray}
\bc{q}&=&\frac{1}{\sqrt{N}}\sum_{\mathbf{k}}\cc{2}{k+q}\ca{1}{k}
\end{eqnarray}
with $N$ the total number of electrons in the doped quantum well. 
In the ground state, the two-dimensional electron gas fills the fundamental subband for $k<k_F$, being $k_F$ the Fermi wave vector.
$\bc{q}$ creates a bright intersubband excitation with in-plane wave vector $\mathbf{q}$, obtained when the two-dimensional electron gas absorbs one cavity photon, as it can be deduced from the light-matter coupling in Eq. (\ref{H}). 

We are interested in calculating the polariton scattering rate induced by the emission of an optical phonon from an initial polariton 'pump' mode (branch $\eta'$ and in-plane wave vector $\textbf{q'}$) to a final 'signal' mode (branch $\eta$ and in-plane wave vector $\textbf{q}$). This kind of process is pictured in Fig. \ref{fig1} for the case $\eta' = UP$ and $\eta = LP$. It is clear that in order to have a sizeable polariton-phonon interaction, both the initial and final polariton modes must have significant electronic components, quantified respectively by $\lvert \be{UP}{q'} \lvert^2$ and $\lvert\be{LP}{q}\lvert^2$. At the same time, in order to have a good coupling to the extracavity electromagnetic field (required for pumping and detection) also the photonic components $\lvert \al{UP}{q'} \lvert^2$ and $\lvert\al{LP}{q}\lvert^2$  need to be significant. These conditions can be simply met when the polariton energy splitting $2\hbar \chi(q_{res}) \sqrt{N}$  at the resonant wavector $q_{res}$ (such as $ \omega_c(q_{res})=\omega_{12}$) is a non negligeable fraction of the optical phonon energy ($36$meV for GaAs). This situation is already realized in recent microcavity samples\cite{Aji_APL,Aji_excited,Luca_PRL} with mid-infrared intersubband transition frequencies. 

If we wish to investigate the occurrence of stimulated scattering, we need to evaluate the scattering rates  for arbitrary occupation numbers $m$ and $n$ of respectively the initial and final polariton modes. 
The emission of an optical phonon can induce the scattering of one polariton from the pump to the signal mode, leading to a transition from the state $ \pc[m]{\eta'}{q'}  \pc[n]{\eta}{q} \lvert F \rangle$ to the state $\dc{q'-q}{q_z} \pc[m-1]{\eta'}{q'}  \pc[n+1]{\eta}{q} \lvert F  \rangle$ where $\lvert F \rangle$ is the ground state of the system (the electronic ground state times the photon and phonon vacuum).
Therefore, we need to consider the squared normalized matrix element $\hbar^2\lvert V^n_m\rvert^{2}$ given by 
\begin{eqnarray}
\label{Vmn}
\frac{\lvert \bra{F}\pa[n+1]{\eta}{q} \pa[m-1]{\eta'}{q'} d_{\mathbf{q'}-\mathbf{q},q_z} H_{phon} \pc[m]{\eta'}{q'} \pc[n]{\eta}{q} \ket{F} \lvert ^2 }{
\bra{F} \pa[n]{\eta}{q}\pa[m]{\eta'}{q'} \pc[m]{\eta'}{q'}\pc[n]{\eta}{q}\ket{F} \bra{F} \pa[n+1]{\eta}{q}\pa[m-1]{\eta'}{q'}  \pc[m-1]{\eta}{q'}\pc[n+1]{\eta'}{q}\ket{F}}.
\end{eqnarray}
 
In order to evaluate Eq. (\ref{Vmn}), we have to exploit the expression of the polariton operators in Eq. (\ref{pol}) in terms of the cavity photon and intersubband excitation operators. To evaluate the matrix elements, we need to commute the destruction operators multiple times to the right side and exploit
the annihilation identity $a_{\mathbf{q}}\lvert F> = b_{\mathbf{q}}\lvert F> = 0$. The cavity photons are elementary bosons obeying the standard commutation rule $[a_{\mathbf{q}}, a^\dagger_{\mathbf{q'}}] = \delta_{\mathbf{q},\mathbf{q'}}$. Instead, intersubband excitation operators are not elementary bosons and satisfy modified commutation rules. We have found that 
\begin{eqnarray}
\comm{\ba{q}}{\bc{q'}}&=&\delta_{\mathbf{q,q'}} -\D{q}{q'}, \\
\D{q}{q'}&=&\delta_{\mathbf{q,q'}} -  \frac{1}{N}\sumk{\lvert \mathbf{k}\lvert < k_F}\cc{1}{k}\ca{1}{k+q-q'}-\cc{2}{k+q'}\ca{2}{k+q} \nonumber,
\end{eqnarray}
where $\D{q}{q'}$ is the operator describing the deviation from the behavior of elementary bosons,
originally introduced in the context of excitonic composite bosons\cite{Combescot1}. 
By iteration, we have found  the following commutation relations
\begin{eqnarray}
\label{multiplecomm}
&\comm{\D{q}{q'}}{\bc[m]{q''}}=\frac{2m}{N}\bc{q''+q'-q}\bc[m-1]{q''}, \\
&\comm{\ba{q}}{\bc[m]{q'}}=m\bc[m-1]{q'}(\delta_{\mathbf{q,q'}}-\D{q}{q'})-\frac{m(m-1)}{N}\bc{2\mathbf{q'-q}}\bc[m-2]{q'}, \nonumber\\
&\comm{\ba[m]{q}}{\bc{q'}}=m(\delta_{\mathbf{q,q'}}-\D{q}{q'})\ba[m-1]{q}-\frac{m(m-1)}{N}\ba{2\mathbf{q-q'}}\ba[m-2]{q}.  \nonumber
\end{eqnarray}
Due to the fact that typical photonic wave vectors $q$ are much smaller (at least two orders of magnitude) than the Fermi wave vector $k_F$, we have  $\D{q}{q'} \lvert F\rangle \simeq 0$ with corrections of the order of $\lvert \mathbf{q-q'} \lvert k_F$ due to the electrons occupying the edge of the Fermi sphere in the ground state. Neglecting these corrections is the only approximation we will assume, which becomes exact in the limit $\lvert \mathbf{q-q'} \lvert  k_F \to 0$.
Exploiting Eq. (\ref{multiplecomm}) some algebra shows that the unnormalized polaritonic matrix element $\bra{F} \pa[n+1]{\eta}{q} \pa[m-1]{\eta'}{q'} d_{\mathbf{q}-\mathbf{q'},q_z} H_{phon} \pc[m]{\eta'}{q'} \pc[n]{\eta}{q} \ket{F}$ is given by \begin{widetext}
\begin{eqnarray}
\label{matrixH}
(n+1)! m!  \be{\eta}{q} \bec{\eta'}{q'} (C_{22,\mathbf{q-q'},q_z}-C_{11,\mathbf{q-q'},q_z})
  \sum_{\substack{l=0,\dots,n \\h=0,\dots,m-1}}  \binom{n}{l} \binom{m-1}{h}  \lvert \al{\eta}{q}\lvert^{2l} \lvert \be{\eta}{q}\lvert^{2(n-l)} 
  \lvert \al{\eta'}{q'}\lvert^{2h}\lvert \be{\eta'}{q'}\lvert^{2(m-1-h)} f^{n-l}_{m-h},
\end{eqnarray}

where
$\label{f}
f_m^n = 
\frac{n}{m}\K{m-1}{n+1}{m}{n-1}+\K{m-1}{n+1}{m-1}{n}
$and the quantity $\K{n}{m}{s}{r}$ is defined by the relation 
\begin{equation}
\label{Kdef}
n! m! \K{n}{m}{s}{r}=\bra{F} \ba[n]{q} \ba[m]{q'}  \bc[s]{q}  \bc[r]{q'} \bc{Q} \ket{F}
\end{equation}
with $\mathbf{Q}=\mathbf{q}(n-s)+\mathbf{q'}(m-r)$. Analogously for the normalization factors in Eq. (\ref{Vmn}), we find
\begin{eqnarray}
\label{normaliz}
\bra{F} \pa[n]{\eta}{q}\pa[m]{\eta'}{q'}\pc[m]{\eta'}{q'} \pc[n]{\eta}{q} \ket{F}&=&
 n!m!\sum_{\substack{l=0,\dots,n \\h=0,\dots,m}} \binom{n}{l} \binom{m}{h} \lvert \al{\eta}{q}\lvert^{2l} \lvert \be{\eta}{q}\lvert ^{2(n-l)} 
 \lvert \al{\eta'}{q'}\lvert^{2h} \lvert \be{\eta'}{q'}\lvert ^{2(m-h)} 
\K{n-l}{m-h}{n-l}{m-h-1}.
 \end{eqnarray}
 \end{widetext}
It is clear that  $\K{n}{m}{s}{r}\propto \delta_{n+m,r+s+1}$ and that 
$\K{n}{m}{n-1}{m}=\K{n}{m}{n}{m-1} $ and $\K{n}{m}{s}{r}=\K{m}{n}{r}{s} $. Using the commutation relations in Eq. (\ref{multiplecomm}) we obtain the the recurrence relation
\begin{eqnarray*}
\label{recurrence}
\begin{array}{ll}
\K{n}{m}{s}{r}=
\delta_{m,r}\delta_{n,s+1}\K{n-1}{m}{n-1}{m-1} +\delta_{m,r+1}\delta_{n,s}\K{n}{m-1}{n-1}{m-1} \\ \\
-\frac{s!r!}{n!m!N} \lbrack n(n-1)\K{s}{r}{n-2}{m} + m(m-1)\K{s}{r}{n}{m-2} +2nm\K{s}{r}{n-1}{m-1}\rbrack
\end{array}
\end{eqnarray*}
that allows us to numerically evaluate the  $\K{m}{n}{s}{r}$ for arbitrary and realistic occupation numbers. Hence, Eq. (\ref{Vmn}) can be rewritten in the simpler form
\begin{eqnarray*}
\lvert V^n_m \lvert^2 =  
 (n+1) m B^n_m \lvert \be{\eta}{q}\be{\eta'}{q'}(C_{22,{\lvert \mathbf{q-q'}\lvert,q_z}} 
 - C_{11,\lvert \mathbf{q-q'}\lvert,q_z}) \lvert ^2 
\end{eqnarray*}
where $B^n_m$ is a bosonicity factor depending on  the coefficients  $\K{n}{m}{s}{r}$.  Its expression is cumbersome, but it can be simply obtained putting together Eqs. (\ref{Vmn}), (\ref{matrixH}) and (\ref{normaliz}). Such a quantity depends on the Hopfield coefficients and on excitation numbers $m$ and $n$ normalized to the total
number of electrons $N$ in the ground state. In the inset of Fig. 2, we report $B^0_m$ versus $m/N$ obtained by a numerical evaluation of our exact recursive relations. For normalized excitation densities $\frac{m+n}{N}$ smaller than $0.1$, we find that   $B^n_m$ is well approximated by the formula
\begin{equation}
B^n_m \simeq 1-\zeta\frac{m+n}{N},
\end{equation}
where $\zeta$ depends on the Hopfield coefficients of the polariton modes and varies from $0$ for pure photonic excitations to $1$ for pure matter ones. Hence, we see that final-state simulation, due to the enhancement factor $(n+1)$, can be obtained when the density of polaritons $(m+n)/S$ is much smaller than the density $N/S$ of the two-dimensional electron gas,  not a stringent condition as shown later.  Indeed, the bosonicity of intersubband cavity polaritons can be controlled by the changing the density of the two-dimensional electron gas, which can be tuned by an electric gate \cite{Aji_APL}. Using Fermi golden rule we find the following formula for the scattering rate
\begin{figure}
\begin{center}
\includegraphics[width=9.6cm]{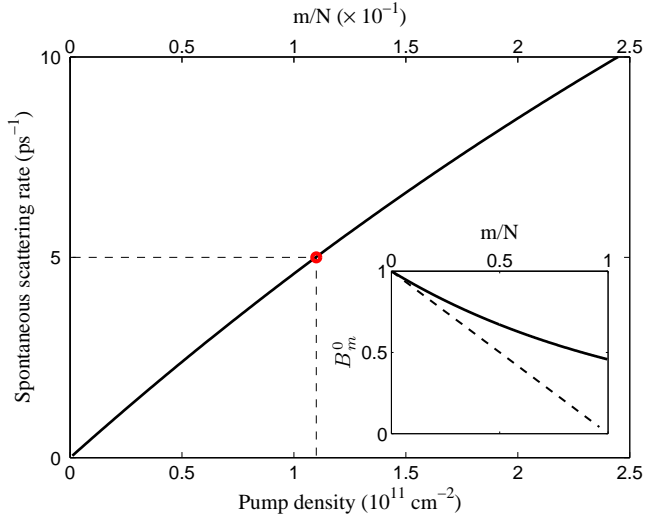}
\caption{\label{fig2}Spontaneous scattering rate $\Gamma^{m,n=0}_{sc}$ for the process depicted in Fig. 1 versus the pump polariton density $m/S$. The electron density in the ground state is $N/S = 10^{12}$cm$^{-2}$.  In the considered range of excitation densities, 
$m/N < 0.25$, i.e., much smaller than the onset of electronic population inversion.  Other GaAs parameters are given in the text. Inset: the solid line represents the bosonicity factor $B^{n=0}_m$ versus $m/N$ for the pump and signal polariton modes considered in Fig. 1. For elementary bosons $B^{n}_m$ is always $1$. The dashed line is the same quantity for pure matter excitations. For $m/N \ll 1$, deviations from perfect bosonicity are negligible.  
For $n \ll m$ (signal much smaller than pump), $B^{n}_m \simeq  B^{n=0}_m$ and the stimulated scattering rate
is $\Gamma^{m,n}_{sc} \simeq (1+n) \Gamma^{m,n=0}_{sc}$ . }
\end{center}
\end{figure}
\begin{eqnarray*}
\Gamma_{sc}^{m,n}=2\pi\sum_{q_z}\int d\omega \lvert V^n_m \lvert^{2}A_{\mathbf{q-q'},q_z,\omega}
\delta(\omega_{\eta,\mathbf{q}}-\omega_{\eta',\mathbf{q'}}+\omega),
\end{eqnarray*}
where $A_{\mathbf{q-q'},q_z,\omega}$ is the spectral function of the optical phonon. Using a Lorentzian shape of width $\Gamma_{LO}$ and neglecting the 
LO-phonon dispersion,  we obtain
\begin{eqnarray}
\label{Gamma}
\Gamma_{sc}^{m,n}= (n+1) B^n_m \lvert \be{\eta}{q}\lvert^2 \lvert \be{\eta'}{q'}\lvert^2 \frac{m}{S}   \frac{\omega_{LO}}{\Gamma_{LO}}\frac{4 e^2 L_{QW} F_{\sigma}}{\epsilon\hbar} ,    
\end{eqnarray}
where $S$ is the sample surface, $L_{QW}$ the quantum well length and $F_{\sigma}$ a form factor (depending on $\sigma=L_{QW}\lvert\mathbf{q-q'}\lvert$) describing the overlap between the conduction subband envelope functions and the optical phonon wavefunction\cite{Bastard}. For typical quantum well widths and photonic wave vectors, $\sigma \ll 1$. In the case of a quantum well with infinite barriers,  $F_{\sigma \simeq 0} \simeq 0.1$. For GaAs optical phonons, the ratio $ \frac{\omega_{LO}}{\Gamma_{LO}}  \approx 100$ \cite{Yu}. In Fig. \ref{fig2}, we report the calculation of the spontaneous scattering rate $\Gamma_{sc}^{m,0}$ (i.e., $n = 0$, unoccupied final state) for the process showed in Fig. \ref{fig1} for a GaAs system with $\hbar \omega_{12} = 150$meV (mid infrared) and  $N/S = 10^{12}$ cm$^{-2}$. The scattering rate into the final signal mode for pump excitation densities around $10^{11}$ cm$^{-2}$ is comparable to the typical loss rate of intersubband cavity polariton modes. Hence, for these pump excitation densities one expects to enter the regime of stimulated scattering. 

Neglecting the pump depletion (relevant only above an eventual stimulation threshold), we can write two rate equations for the signal and pump populations, namely
\begin{equation}
\label{popequation}
\frac{dn}{dt}=\Gamma_{sc}^{m,n}-\Gamma_{loss}n;~~
\frac{dm}{dt} =\frac{A I_{pump} S}{\hbar \omega_{UP,q'}}-\Gamma'_{loss} m,
\end{equation}
where $\Gamma_{loss}$ and $\Gamma'_{loss}$ are the loss rates of the signal and pump modes, $A$ the polariton absorption coefficient at the pump frequency and $I_{pump}$ the optical pump intensity.
From the steady-state solution for $n$, we can calculate the threshold pump density $m_{thr}/S$ to have a lasing instability. For $n \ll m$, $B^n_m \simeq B^0_m$ and  $\Gamma_{sc}^{m,n} \simeq (1+n)  \Gamma_{sc}^{m,0}$. The threshold pump polariton density $m_{thr}/S$ is then given by the equation $\Gamma_{sc}^{m_{thr},0}=\Gamma_{loss}$. The steady-state solution for $m$ gives the threshold pumping intensity versus the polariton threshold density, namely
$
I^{thr}_{pump}=\frac{\Gamma'_{loss}  \hbar \omega_{UP,q'}}{A} m_{thr}/S$.
For a realistic $\Gamma_{loss}=5$ ps$^{-1}$, we obtain a threshold  density for the pump mode of $1.1 \times10^{11}$cm$^{-2}$, i.e. $m/N = 0.11$, as indicated in Fig. 2. With a polariton absorption coefficient $A=0.4$ \cite{Luca_PRL}, this gives an impressively low threshold pump intensity of $3.5\times 10^{4}W/$cm$^2$. This is approximately 2 orders of magnitude smaller of what required to achieve electron population inversion in the two subbands\cite{Julien97}. 


In conclusion, we have derived a theory for the stimulated scattering of intersubband cavity
polariton excitations of a dense two-dimensional electron gas. The intersubband cavity polariton excitations are composite bosons arising from the strong light-matter coupling and are not associated to any bound electronic states. We have shown exactly how the bosonicity of these excitations is controlled by density of the two-dimensional electron gas in the ground state.  Our exact results show that ultralow threshold intersubband cavity polariton lasing without population inversion can be achieved under optical pumping by exploiting the efficient interaction with optical phonons. The present theory could pave the way to the experimental demonstration of fundamental quantum degeneracy phenomena and unconventional lasing devices based on composite bosons with controllable properties and interactions.

We wish to thank  S. Barbieri, R. Colombelli, F. Julien and  C. Sirtori for discussions.

\end{document}